\documentclass[aps,prd,preprint,showpacs]{revtex4-1}
\usepackage{amsmath,amssymb}
\usepackage{indentfirst,latexsym,bm}
\begin{document}
\title{CP and CPT Violating Parameters Determined from the Joint Decays of $C=+1$ Entangled Neutral Pseudoscalar Mesons}
\author{Zhijie Huang}
\affiliation{Center for Field Theory and Particle Physics, Department of Physics, Fudan University, Shanghai, 200433, China}
\author{Yu Shi}
\email{Corresponding author, email: yushi@fudan.edu.cn}
\affiliation{Center for Field Theory and Particle Physics, Department of Physics, Fudan University, Shanghai, 200433, China}
\begin{abstract}
Entangled pseudoscalar neutral meson pairs have been used in studying CP violation and searching CPT violation, but almost all the previous works concern $C=-1$ entangled state. Here we consider $C=+1$ entangled state of pseudoscalar neutral mesons, which is quite different from $C=-1$ entangled state and provides complementary information on symmetry violating parameters.  After developing a general formalism, we consider three kinds of decay processes, namely, semileptonic-semileptonic, hadronic-hadronic and semileptonic-hadronic processes. For each kind of processes, we calculate the integrated rates  of joint decays with a fixed time interval, as well as asymmetries defined for these joint rates of different channels.  In turn, these asymmetries can be used to determine the four real numbers of the two indirect symmetry violating parameters, based on a general relation between the symmetry violating parameters and the decay asymmetries presented here.  Various discussions are made on indirect and direct violations and  the violation of $\Delta {\cal F} =\Delta Q$ rule, with some results presented as theorems.
\end{abstract}
\pacs{11.30.Er, 13.20.-v, 13.25.-k}
\maketitle

\section{introduction}

Pseudoscalar neutral mesons are important in studying CP violation, as well as CPT violation, which is important in the standard model extension~\cite{kos1}. Moreover, these particles in entangled or EPR correlated states have also been used in discussing these violations~\cite{lee,inglis,day,lipkin}. Several experimental groups have investigated the violations of the CP and CPT symmetries in the entangled  pseudoscalar neutral mesons, such as $B_d\bar{B}_d$ pairs produced in $\Upsilon(4\textbf{S})$ resonance and $B_s\bar{B}_s$ pairs produced in $\Upsilon(5\textbf{S})$ resonance~\cite{aubert1,aubert2,abe,aubert3,aubert4,louvot,aqui,lees}, as well as $K^0\bar{K}^0$ pairs produced in $\phi$ resonance~\cite{ambro,di1,di2,ame}. Various theoretical studies have also been made~\cite{bernabeu1,dunietz,buchanan,dambrosio,bernabeu,
kobayashi,kos2,bernabeu2,baji,petrov,bigi,soni,kittle,yu1,yu2}. However, most of them concern $C=-1$ state $|\Psi_-\rangle$.

On the other hand, it has been known that $C=+1$ entangled state $|\Psi_+\rangle$ can also be produced, for example,  for $B_s\bar{B}_s$ pairs produced in  the $\Upsilon(5\textbf{S})$ resonance with  $10\%$ branch ratio~\cite{soni,louvot,aqui} and, most remarkably,  for $B_d\bar{B}_d$ pairs in an energy range just  above the  $\Upsilon(4\textbf{S})$ resonance with 100\% branch ratio~\cite{akerib}. Hence it is very  interesting to investigate the decay properties of the $C=+1$ entangled state, which is the purpose of this paper. Towards the end of the paper, we shall note some complementarities between the uses of $|\Psi_+\rangle$ and $|\Psi_-\rangle$.

After a review of various CP and CPT violating parameters and the relations among them in Sec.~\ref{evolution},  we calculate the  rates of the joint decays of  the $C=+1$ entangled meson pairs in Sec.~\ref{jointdecays}. Then we discuss various  experimentally observable  asymmetries between different joint decay rates in Sec.~\ref{asymmetries}, giving the general expressions in Subsection \ref{general}, and  considering  the semileptonic-semileptonic decays  in Subsection \ref{semilepton},  the     hadronic-hadronic decays in Subsection \ref{hadron},  and the    semileptonic-hadronic decays in Subsection \ref{semihadr}. Subsequently in Sec.~\ref{calculation}, we discuss how to obtain the four  real numbers of CP and CPT symmetry violating parameters from  the asymmetries of joint decays.  In Sec.~\ref{discussions}, we discuss some specific experimentally relevant cases, and present some simple results in the form of theorems.  A summary is made in Sec.~\ref{summary}.

\section{ Indirect symmetry violating parameters and time evolution \label{evolution}}

As usual, we denote the  pseudoscalar neutral meson with the flavor eigenvalue $+1$ as $|M^0\rangle$, and  its antiparticle  with the flavor eigenvalue $-1$  as  $|\bar{M}^0\rangle \equiv CP|M^0\rangle$. The time-dependent state of a single meson is
\begin{equation}
|M(t)\rangle=\alpha(t)|M^0\rangle+\beta(t)|\bar{M}^0\rangle,
\end{equation}
where $\alpha(t)$ and $\beta(t)$ are determined by
\begin{equation}
i\frac{d}{dt}{\alpha(t) \choose \beta(t)}=
\left( \begin{array}{cc}
H_{11} & H_{12} \\
H_{21} & H_{22}
\end{array}\right){\alpha(t) \choose \beta(t)}.
\end{equation}
The effective Hamiltonian $H$ has the following properties,
\begin{itemize}
  \item  if CPT or CP is conserved, then $H_{11}=H_{22}$,
  \item  if $T$ or CP is conserved, then $H_{12}=H_{21}$.
\end{itemize}

One can define~\cite{chou}
\begin{equation}
\begin{split}
\delta_M\equiv&\frac{H_{22}-H_{11}}{\sqrt{H_{12}H_{21}}},\\
\epsilon_M\equiv&\frac{\sqrt{H_{12}}-\sqrt{H_{21}}}{\sqrt{H_{12}}+\sqrt{H_{21}}}.
\end{split}
\label{para1}
\end{equation}
Indirect CPT conservation implies $\delta_M=0$, while indirect CP conservation implies $\epsilon_M=0$ and $\delta_M=0$.

The eigenvalues of $H$ are
\begin{equation}
\begin{split}
\lambda_L\equiv m_L-\frac{i}{2}\Gamma_L=\frac{1}{2}\Big[H_{11}+H_{22}-\sqrt{(H_{11}-H_{22})^2+4H_{12}H_{21}}\Big],\\
\lambda_S\equiv m_S-\frac{i}{2}\Gamma_S=\frac{1}{2}\Big[H_{11}+H_{22}+\sqrt{(H_{11}-H_{22})^2+4H_{12}H_{21}}\Big],
\end{split}
\end{equation}
and the corresponding eigenstates are
\begin{equation}
\begin{split}
|M_L\rangle=\frac{1}{\sqrt{|p_L|^2+|q_L|^2}}\big(p_L|M^0\rangle-q_L|\bar{M}^0\rangle\big),\\
|M_S\rangle=\frac{1}{\sqrt{|p_S|^2+|q_S|^2}}\big(p_S|M^0\rangle+q_S|\bar{M}^0\rangle\big).
\end{split}
\end{equation}
Defining
\begin{equation}
\frac{1+\Delta_M}{1-\Delta_M}\equiv\frac{\delta_M}{2}+\sqrt{1+\frac{\delta_M^2}{4}}, \label{ded}
\end{equation}
we have
\begin{equation*}
\begin{split}
\frac{p_L}{q_L}=&\frac{(1+\epsilon_M)(1+\Delta_M)}{(1-\epsilon_M)(1-\Delta_M)},\\
\frac{p_S}{q_S}=&\frac{(1+\epsilon_M)(1-\Delta_M)}{(1-\epsilon_M)(1+\Delta_M)}.
\end{split}
\label{xishu}
\end{equation*}

One also defines
\begin{equation}
\begin{split}
\frac{p_L}{q_L}&\equiv\frac{1+\epsilon_L}{1-\epsilon_L},\\
\frac{p_S}{q_S}&\equiv\frac{1+\epsilon_S}{1-\epsilon_S},
\end{split}
\end{equation}
and
\begin{equation}
\begin{split}
\delta=&\frac{1}{2}(\epsilon_S-\epsilon_L)=\frac{H_{11}-H_{22}}{H_{12}+H_{21}+\sqrt{(H_{11}-H_{22})^2+4H_{12}H_{21}}},\\
\epsilon=&\frac{1}{2}(\epsilon_S+\epsilon_L)=\frac{H_{12}-H_{21}}{H_{12}+H_{21}+\sqrt{(H_{11}-H_{22})^2+4H_{12}H_{21}}}.
 \end{split}
\end{equation}
Hence $\delta=0$ corresponds to $H_{11}=H_{22}$ while $\epsilon=0$ corresponds to $H_{12}=H_{21}$.
($\delta_M, \epsilon_M$) and ($\delta, \epsilon$) are related as
\begin{equation}
\begin{split}
\delta=&-\frac{1}{2}\frac{\delta_M(1-\epsilon_M^2)}{1+\epsilon_M^2+(1-\epsilon_M^2)\sqrt{1+\frac{\delta_M^2}{4}}},\\
\epsilon=&
\frac{2\epsilon_M}{1+\epsilon_M^2+(1-\epsilon_M^2)
\sqrt{1+\frac{\delta_M^2}{4}}}.
\end{split}
\label{relat}
\end{equation}
We would like to emphasize that  $\delta$ and $\epsilon$ are each  dependent on both $\delta_M$ and $\epsilon_M$, that is, a nonzero value of $\delta$ or $\epsilon$  corresponds to mixing of CP and CPT violations.  Moreover, as seen in (\ref{ded}), $\delta_M=0$ is equivalent to $\Delta_M=0$, and using $\Delta_M$ can avoid square roots in the calculations.

Therefore,  in the following, we use the parameters  ($\Delta_M$,$\epsilon_M$) in characterizing indirect symmetry violations. $\epsilon_M \neq 0$ implies indirect CP violation, while $\Delta_M \neq 0$ implies indirect CPT violation and indirect CP violation.

With $|M^0(t=0)\rangle\equiv|M^0\rangle$ and $|\bar{M}^0(t=0)\rangle\equiv|\bar{M}^0\rangle$, we have
\begin{eqnarray}
|M^0(t)\rangle&=&\frac{1}{2}\big[(1-\xi)e^{-i\lambda_St}+
(1+\xi)e^{-i\lambda_Lt}\big]|M^0\rangle+
\frac{1}{2}\eta_1(e^{-i\lambda_St}-e^{-i\lambda_Lt})|\bar{M}^0\rangle, \label{m0t} \\
|\bar{M}^0(t)\rangle&=&\frac{1}{2}\eta_2(e^{-i\lambda_St}-e^{-i\lambda_Lt})
|M^0\rangle+\frac{1}{2}\big[(1+\xi)e^{-i\lambda_St}+(1-\xi)e^{-i\lambda_Lt}
\big]|\bar{M}^0\rangle,
\label{m00t}
\end{eqnarray}
where
\begin{equation}
\begin{split}
\xi& \equiv \frac{2\Delta_M}{1+\Delta_M^2},\\
\eta_1& \equiv \frac{(1-\epsilon_M)(1-\Delta_M^2)}{(1+\epsilon_M)(1+\Delta_M^2)},\\
\eta_2& \equiv \frac{(1+\epsilon_M)(1-\Delta_M^2)}{(1-\epsilon_M)(1+\Delta_M^2)}.
\end{split}
\label{xieta}
\end{equation}

Now we consider the $C=+1$ entangled state, shared by particles $a$ and $b$,
\begin{equation}
|\Psi_+\rangle=\frac{1}{\sqrt{2}}
\big[|M^0\rangle|\bar{M}^0\rangle+|\bar{M}^0\rangle|M^0\rangle\big].
\label{psi00}
\end{equation}
where  each term in the form of $|x\rangle|y\rangle$ apparently means a direct product of $|x\rangle$ of a particle and $|y\rangle$ of b particle.

The joint probability, or the joint decay rate,  that particle $a$ decays to $\psi^a$ at time $t_a$ while particle $b$  decays to $\psi^b$ at time $t_b$ can be obtained as
\begin{equation}
I(\psi^a,t_a;\psi^b,t_b)=|\langle \psi^a\psi^b|{\cal H}_a{\cal H}_b|\Psi(t_a,t_b)\rangle|^2,
\label{generaldecay}
\end{equation}
where  ${\cal H}_{\alpha}$ is the weak interaction Hamiltonian governing the decay of particle $\alpha=a,b$, the time-dependent state  $|\Psi_+(t_a,t_b)\rangle$ is given by
\begin{equation}
|\Psi_+(t_a,t_b)\rangle=
\frac{1}{\sqrt{2}}
\big[|M^0(t_a)\rangle|\bar{M}^0(t_b)\rangle+|\bar{M}^0(t_a)\rangle|M^0(t_b)
\rangle\big],
\end{equation}
where $|M^0(t_\alpha)\rangle$ and  $|\bar{M}^0(t_\alpha)\rangle$, with $\alpha=a,b$, are as given in Equations (\ref{m0t}) and (\ref{m00t}). This standard treatment using the decay times $t_a$ and $t_b$ of the two entangled mesons~\cite{inglis,day,lipkin} gives the same result as that of the approach taking account of the two measurements at  $t_a$ and $t_b$.

By substituting $|M^0(t_\alpha)\rangle$ and  $|\bar{M}^0(t_\alpha)\rangle$, one obtains
\begin{equation}
\begin{split}
|\Psi_+(t_a,t_b)\rangle=&\frac{1}{2\sqrt{2}}
\Big\{
\big[\eta_2(1-\xi)e^{-i\lambda_S(t_a+t_b)}
+\eta_2\xi e^{-i(\lambda_St_a+\lambda_Lt_b)}
+\eta_2\xi e^{-i(\lambda_Lt_a+\lambda_St_b)}\\
&-\eta_2(1+\xi)e^{-i\lambda_L(t_a+t_b)}\big]|M^0M^0\rangle\\
&+\big[(1-\xi^2)e^{-i\lambda_S(t_a+t_b)}
-\xi(1-\xi) e^{-i(\lambda_St_a+\lambda_Lt_b)}\\
& +\xi (1+\xi)  e^{-i(\lambda_Lt_a+\lambda_St_b)}
+(1-\xi^2 )e^{-i\lambda_L(t_a+t_b)}\big]|M^0\bar{M}^0\rangle\\
&+\big[(1-\xi^2 )e^{-i\lambda_S(t_a+t_b)}
+\xi(1+\xi) e^{-i(\lambda_St_a+\lambda_Lt_b)}\\
& -\xi (1-\xi)  e^{-i(\lambda_Lt_a+\lambda_St_b)}
+(1-\xi^2  )e^{-i\lambda_L(t_a+t_b)}
  \big]|\bar{M}^0M^0\rangle\\
&+\big[\eta_1(1+\xi)e^{-i\lambda_S(t_a+t_b)}
-\eta_1\xi e^{-i(\lambda_St_a+\lambda_Lt_b)}\\
& -\eta_1\xi e^{-i(\lambda_Lt_a+\lambda_St_b)}
-\eta_1(1-\xi)e^{-i\lambda_L(t_a+t_b)}
\big] |\bar{M}^0\bar{M}^0\rangle \Big\}.
\end{split}
\label{psitt}
\end{equation}

Note that the initial entangled state $|\Psi_+\rangle$ can  be rewritten in CP basis {\em exactly}  as
\begin{equation}
|\Psi_+\rangle=\frac{1}{\sqrt{2}}\big[|M_+\rangle|M_+\rangle
-|M_-\rangle|M_-\rangle\big],
\label{psi+-}
\end{equation}
where
\begin{equation*}
\begin{split}
|M_\pm \rangle=\frac{1}{\sqrt{2}}(|M^0\rangle  \pm |\bar{M}^0\rangle)
\end{split}
\end{equation*}
is CP eigenstate of eigenvalue $\pm 1$. If needed, $|\Psi_+(t_a,t_b)\rangle$ can also be rewritten as
\begin{equation}
|\Psi_+(t_a,t_b)\rangle=\frac{1}{\sqrt{2}}\big[|M_+(t_a)\rangle|M_+(t_b)\rangle
-|M_-(t_a)\rangle|M_-(t_b)\rangle\big],
\end{equation}
where
\begin{equation}
\begin{array}{cl}
|M_+(t)\rangle=&\frac{1}{4}\big\{[(2+\eta_1+\eta_2)e^{-i\lambda_St}+(2-\eta_1-\eta_2)e^{-i\lambda_Lt}]|M_+\rangle\\
&-(2\xi-\eta_2+\eta_1)(e^{-i\lambda_St}-e^{-i\lambda_Lt})|M_-\rangle\big\},\\
|M_-(t)\rangle=&\frac{1}{4}\big\{[(2-\eta_1-\eta_2)e^{-i\lambda_St}+(2+\eta_1+\eta_2)e^{-i\lambda_Lt}]|M_-\rangle\\
&-(2\xi+\eta_2-\eta_1)(e^{-i\lambda_St}-e^{-i\lambda_Lt})|M_+\rangle\big\}.
\end{array}
\label{m+-t}
\end{equation}

In comparison,  $C=-1$ state is
\begin{equation}
|\Psi_-\rangle=\frac{1}{\sqrt{2}}
\big[|M^0\rangle|\bar{M}^0\rangle-|\bar{M}^0\rangle|M^0\rangle\big],
\end{equation}
which can be rewritten in CP basis {\em exactly} as
\begin{eqnarray}
|\Psi_-\rangle &=& \frac{1}{\sqrt{2}}(|M_-\rangle|M_+\rangle-
|M_+\rangle|M_-\rangle).
\end{eqnarray}
Note that in CP basis, $|\Psi_+\rangle$ is a superposition of equal-CP products  $|M_+\rangle|M_+\rangle$
and $|M_-\rangle|M_-\rangle$, while $|\Psi_-\rangle$ is a superposition of unequal-CP products  $|M_-\rangle|M_+\rangle$ and $|M_+\rangle|M_-\rangle$.

The time-dependent state  $|\Psi_-(t_a,t_b)\rangle$ is given by
\begin{equation}
\begin{split}
|\Psi_-(t_a,t_b)\rangle=&
\frac{1}{\sqrt{2}}
\big[|M^0(t_a)\rangle|\bar{M}^0(t_b)\rangle-|\bar{M}^0(t_a)\rangle|M^0(t_b)
\rangle\big],\\
=&\frac{1}{2\sqrt{2}}
\big\{-\eta_2(e^{-i(\lambda_St_a+\lambda_Lt_b)}
-e^{-i(\lambda_Lt_a+\lambda_St_b)})|M^0M^0\rangle\\
&+[(1-\xi)e^{-i(\lambda_St_a+\lambda_Lt_b)}
+(1+\xi)e^{-i(\lambda_Lt_a+\lambda_St_b)}]|M^0\bar{M}^0\rangle\\
&-[(1+\xi)e^{-i(\lambda_St_a+\lambda_Lt_b)}+
(1-\xi)e^{-i(\lambda_Lt_a+\lambda_St_b)}]|\bar{M}^0 M^0\rangle\\
&+\eta_1(e^{-i(\lambda_St_a+\lambda_Lt_b)}-
e^{-i(\lambda_Lt_a+\lambda_St_b)})|\bar{M}^0\bar{M}^0\rangle\big\},
\end{split}
\label{minuspsitt}
\end{equation}
which can  be rewritten as
\begin{eqnarray}
|\Psi_-(t_a,t_b)\rangle &=& \frac{1}{\sqrt{2}}(|M_-(t_a)\rangle|M_+(t_b)\rangle-
|M_+(t_a)\rangle|M_-(t_b)\rangle).
\end{eqnarray}

\section{integrated rates of joint decays \label{jointdecays} }

We denote  the decay amplitudes
\begin{equation}
\begin{split}
r^a\equiv&\langle \psi^a|{\cal H}|M^0\rangle,\\
r^b\equiv&\langle \psi^b|{\cal H}|M^0\rangle,\\
\bar{r}^a\equiv&\langle \psi^a|{\cal H}|\bar{M}^0\rangle,\\
\bar{r}^b\equiv&\langle \psi^b|{\cal H}|\bar{M}^0\rangle.
\end{split}
\label{xi}
\end{equation}

Suppose that particle $a$ decays to $|\psi^a\rangle$ at $t_a$, while particle $b$ decays to $|\psi^b\rangle$ at $t_b$.
The joint decay rate is obtained as
\begin{equation}
\begin{split}
I(\psi^a,t_a;\psi^b,t_b)=&\big|\langle \psi^a\psi^b|{\cal H}_a{\cal H}_b|\Psi_+(t_a,t_b)\big|^2\\
=&\frac{1}{8}\big\{e^{-\Gamma_S(t_a+t_b)}|\Theta|^2
+2e^{-(\Gamma_St_a+\Gamma t_b)}\Re[\Theta^*\Xi e^{-i\Delta mt_b}]+2e^{-(\Gamma t_a+\Gamma_St_b)}\Re[\Theta^*\Phi e^{-i\Delta mt_a}]\\
&+2e^{-\Gamma(t_a+t_b)}\Re[\Theta^*\Lambda e^{-i\Delta m(t_a+t_b)}]+e^{-(\Gamma_St_a+\Gamma_Lt_b)}|\Xi|^2
+2e^{-\Gamma(t_a+t_b)}\Re[\Xi^*\Phi e^{i\Delta m(t_b-t_a)}]\\&
+2e^{-(\Gamma t_a+\Gamma_Lt_b)}\Re[\Xi^*\Lambda  e^{i\Delta mt_a}]
+e^{-(\Gamma_Lt_a+\Gamma_St_b)}|\Phi|^2+2e^{-(\Gamma_Lt_a+\Gamma t_b)}\Re[\Phi^*\Lambda e^{i\Delta mt_b}]\\
&+e^{-\Gamma_L(t_a+t_b)}|\Lambda|^2\big\},
\end{split}
\label{ittgeneral}
\end{equation}
where
\begin{equation}
\begin{split}
\Gamma& \equiv \frac{1}{2}(\Gamma_S+\Gamma_L),\\
\Delta m &\equiv m_L-m_S,\\
\Delta\Gamma &\equiv \Gamma_L-\Gamma_S,\\
\Theta&= \eta_2(1-\xi)r^ar^b
+(1-\xi^2)(r^a\bar{r}^b+r^b\bar{r}^a)
+\eta_1(1+\xi)\bar{r}^a\bar{r}^b,\\
\Xi&= \eta_2\xi r^ar^b
-\xi(1-\xi) r^a\bar{r}^b
+\xi(1+\xi) r^b\bar{r}^a
-\eta_1\xi \bar{r}^a\bar{r}^b, \\
\Phi&=\eta_2\xi r^ar^b
+\xi(1+\xi) r^a\bar{r}^b
-\xi (1-\xi) r^b\bar{r}^a
-\eta_1\xi \bar{r}^a\bar{r}^b, \\
\Lambda&=-\eta_2(1+\xi)r^ar^b
+(1-\xi^2) (r^a\bar{r}^b+r^b\bar{r}^a)
- \eta_1(1-\xi)\bar{r}^a\bar{r}^b.\\
\end{split}
\label{Xi}
\end{equation}

Consider a fixed time interval $\Delta t=t_b-t_a$ between $t_a$ and $t_b$, we obtain the time-integrated decay rate
\begin{equation}
\begin{split}
I'(\psi^a,\psi^b,\Delta t)=&\int^\infty_0dt_aI(\psi^a,t_a;\psi^b,t_a+\Delta t)\\
\equiv &\frac{1}{8}\Big\{\frac{e^{-\Gamma_S\Delta t}}{2\Gamma_S}|\Theta|^2
+2e^{-\Gamma\Delta t}\Re\big[\frac{\Theta^*\Xi e^{-i\Delta m\Delta t}}{\Gamma_S+\Gamma+i\Delta m}\big]+2e^{-\Gamma_S\Delta t}\Re\big[\frac{\Theta^*\Phi}{\Gamma_S+\Gamma+i\Delta m}\big]\\
&+e^{-\Gamma\Delta t}\Re\big[\frac{\Theta^*\Lambda e^{-i\Delta m\Delta t}}{\Gamma+i\Delta m}\big]+\frac{e^{-\Gamma_L\Delta t}}{\Gamma_S+\Gamma_L}|\Xi|^2
+\frac{e^{-\Gamma\Delta t}}{\Gamma}\Re[\Xi^*\Phi e^{i\Delta m\Delta t}]\\
&+2e^{-\Gamma_L\Delta t}\Re\big[\frac{\Xi^*\Lambda}{\Gamma_L+\Gamma-i\Delta m}\big]
+\frac{e^{-\Gamma_S\Delta t}}{\Gamma_S+\Gamma_L}|\Phi|^2+2e^{-\Gamma\Delta t}\Re\big[\frac{\Phi^*\Lambda e^{i\Delta m\Delta t}}{\Gamma_L+\Gamma-i\Delta m}\big]\\
&+\frac{e^{-\Gamma_L\Delta t}}{2\Gamma_L}|\Lambda|^2\Big\}.
\end{split}
\label{iintegral}
\end{equation}

Ignoring the higher orders of   $\Delta_M$  and $\epsilon_M$, we have
\begin{equation}
\begin{split}
\xi&\approx2\Delta_M,\\
\eta_1&\approx1-2\epsilon_M,\\
\eta_2&\approx1+2\epsilon_M,
\end{split}
\label{xietasimeq}
\end{equation}
therefore
\begin{equation}
\begin{array}{cl}
I(\psi^a,t_a;\psi^b,t_b)\approx&\frac{1}{2}\big[\frac{1}{4}
\tilde{f}_0^{r^ar^b}(t_a,t_b)+\tilde{f}_1^{r^ar^b}(t_a,t_b)\Re\Delta_M+\tilde{f}_2^{r^ar^b}(t_a,t_b)\Im\Delta_M\\
&+\tilde{f}_3^{r^ar^b}(t_a,t_b)\Re\epsilon_M
+\tilde{f}_4^{r^ar^b}(t_a,t_b)\Im\epsilon_M\big],
\end{array}
\label{ittsimeq}
\end{equation}
where the functions $\tilde{f}_i^{r^ar^b}$,
$(i=0,1,2,3,4$), are as  given in Appendix. One also obtains the integrated rates
\begin{equation}
\begin{split}
I'(\psi^a,\psi^b,\Delta t)
&\approx\frac{1}{2}\big[\frac{1}{4}f_0^{r^ar^b}(\Delta t)+f_1^{r^ar^b}(\Delta t)\Re\Delta_M+f_2^{r^ar^b}(\Delta t)\Im\Delta_M\\
&+f_3^{r^ar^b}(\Delta t)\Re\epsilon_M
+f_4^{r^ar^b}(\Delta t)\Im\epsilon_M\big],
\end{split}
\label{deltat}
\end{equation}
with
\begin{equation}
f^{r^ar^b}_i(\Delta t) = \int_0^\infty \tilde{f}^{r^ar^b}_i(t_a,t_a+\Delta t) dt_a,
\label{fx}
\end{equation}
the details of which are also given in Appendix.

\section{Asymmetries between different joint decay rates \label{asymmetries} }

\subsection{General formalism \label{general} }

In general, one can consider the asymmetry $\mathcal{A}(\psi^1\psi^2,\psi^3\psi^4;\Delta t)$ between the  rate $I'(\psi^1,\psi^2,\Delta t)$ of the joint decays with $\psi^a=\psi^1$ and $\psi^b=\psi^2$ and the rate $I'(\psi^3,\psi^4,\Delta t)$ with $\psi^a=\psi^3$ and $\psi^b=\psi^4$,
\begin{eqnarray}
\mathcal{A}(\psi^1\psi^2,\psi^3\psi^4;\Delta t)&\equiv& \frac{I'(\psi^1,\psi^2,\Delta t)-I'(\psi^3,\psi^4,\Delta t)}{I'(\psi^1,\psi^2,\Delta t)+I'(\psi^3,\psi^4,\Delta t)}, \label{gen}
\\& \approx& \frac{\frac{1}{4}(f_0^{r^1r^2}-f_0^{r^3r^4})
+\sum_{i=1}^{4}(f_i^{r^1r^2}-f_i^{r^3r^4})\sigma_i}{
\frac{1}{4}(f_0^{r^1r^2}+f_0^{r^3r^4})+
\sum_{i=1}^{4}(f_i^{r^1r^2}+f_i^{r^3r^4})\sigma_i},
\label{index}
\end{eqnarray}
where we have introduced shorthand notations $\sigma_1\equiv\Re{\Delta_M}$, $\sigma_2\equiv\Im{\Delta_M}$, $\sigma_3\equiv\Re{\epsilon_M}$ and $\sigma_4\equiv\Im{\epsilon_M}$.

In particular, we shall study the equal-state asymmetry
\begin{equation}\begin{array}{rcl}
A(\psi^1\psi^1,\psi^2\psi^2;\Delta t) &\equiv & \frac{I'(\psi^1,\psi^1,\Delta t)-I'(\psi^2,\psi^2,\Delta t)}{I'(\psi^1,\psi^1,\Delta t)+I'(\psi^2,\psi^2,\Delta t)} \\
&\approx&\frac{\frac{1}{4}(f_0^{r^1r^1}-f_0^{r^2r^2})+\sum_{i=1}^4(f_i^{r^1r^1}-f_i^{r^2r^2})\sigma_i}{\frac{1}{4}(f_0^{r^1r^1}+f_0^{r^2r^2})+\sum_{i=1}^4(f_i^{r^1r^1}+f_i^{r^2r^2})\sigma_i},
\end{array}
\label{eq}
\end{equation}
and the unequal-state asymmetry
\begin{equation}\begin{array}{rcl}
A(\psi^1\psi^2,\psi^2\psi^1;\Delta t)&\equiv& \frac{I'(\psi^1,\psi^2,\Delta t)-I'(\psi^2,\psi^1,\Delta t)}{I'(\psi^1,\psi^2,\Delta t)+I'(\psi^2,\psi^1,\Delta t)} \label{uneq} \\&\approx&
\frac{\sum_{i=1}^4(f_i^{r^1r^2}-f_i^{r^2r^1})\sigma_i}{\frac{1}{2}f_0^{r^1r^2}+\sum_{i=1}^4(f_i^{r^1r^2}+f_i^{r^2r^1})\sigma_i},
\end{array}
\end{equation}
where we have used the property $f_0^{r^1r^2}=f_0^{r^2r^1}$, which can be seen from the expression of $f_0^{r^ar^b}$ in the Appendix. In the following, we will  discuss three different kinds of processes.

\subsection{Semileptonic-semileptonic Processes} \label{semilepton}

Consider the semileptonic-semileptonic decay processes with the final states $|\psi^1\rangle=|l^+\rangle$ and $|\psi^2\rangle=|l^-\rangle$, which are  flavor eigenstates with eigenvalues $1$ and $-1$, respectively. The decay amplitudes are $\langle l^+|{\cal H}|M^0\rangle\equiv {R^+}$, $\langle l^-|{\cal H}|M^0\rangle\equiv {S^-}$, $\langle l^+|{\cal H}|\bar{M}^0\rangle\equiv {\bar{R}^+}$ and $\langle l^-|{\cal H}|\bar{M}^0\rangle\equiv {\bar{S}^-}$.

From Eq.~(\ref{ittgeneral}), $I(l^+,t_a;l^+,t_b)$ is obtained by substitution  $(r^a,r^b,\bar{r}^a,\bar{r}^b)=(R^+,R^+,\bar{R}^+,\bar{R}^+)$, $I(l^+,t_a;l^-,t_b)$ by substitution $(r^a,r^b,\bar{r}^a,\bar{r}^b)=(R^+,S^-,\bar{R}^+,\bar{S}^-)$, $I(l^-_a,t_a;l^+_b,t_b)$ by substitution $(r^a,r^b,\bar{r}^a,\bar{r}^b)=(S^-,R^+,\bar{S}^-,\bar{R}^+)$ and $I(l^-_a,t_a;l^-_b,t_b)$ by  substitution $(r^a,r^b,\bar{r}^a,\bar{r}^b)=(S^-,S^-,\bar{S}^-,\bar{S}^-)$.

In this case,  with $|\psi^1\rangle = |l^+\rangle$ and $|\psi^2\rangle=|l^-\rangle$, the   equal-flavor asymmetry is
\begin{equation}
\begin{array}{cl}
A(l^+l^+,l^-l^-;\Delta t)
&\approx \frac{\frac{1}{4}(f_0^{R^+R^+}-f_0^{S^-S^-})+\sum_{i=1}^4(f_i^{R^+R^+}-f_i^{S^-S^-})\sigma_i}{\frac{1}{4}(f_0^{R^+R^+}+f_0^{S^-S^-})+\sum_{i=1}^4(f_i^{R^+R^+}+f_i^{S^-S^-})\sigma_i},
\end{array}\label{A1} \\
\end{equation}
while the unequal-flavor asymmetry is
\begin{equation}\begin{array}{cl}
 A(l^+l^-,l^-l^+;\Delta t)
&\approx\frac{\sum_{i=1}^4(f_i^{R^+S^-}-f_i^{S^-R^+})\sigma_i}{\frac{1}{2}f_0^{R^+S^-}+
\sum_{i=1}^4(f_i^{R^+S^-}+f_i^{S^-R^+})\sigma_i}.
\end{array} \label{A2} \\
\end{equation}
They are obtained from Equations (\ref{eq}) and  (\ref{uneq}) respectively, using the substitution
$r^1= {R^+}$, $r^2 = {S^-}$, $\bar{r}^1= {\bar{R}^+}$ and $\bar{r}^2= {\bar{S}^-}$.

\subsection{Hadronic-hadronic Processes}\label{hadron}

For the hadronic-hadronic  processes, we denote the two final states as  $|\psi^1\rangle=|h_1\rangle$ and $|\psi^2\rangle=|h_2\rangle$. The decay amplitudes are
$\langle h_1|{\cal H}|M^0\rangle\equiv Q_1$, $ \langle h_2|{\cal H}|M^0\rangle\equiv Q_2,$ $\langle h_1|{\cal H}|\bar{M}^0\rangle\equiv \bar{Q}_1,$ $\langle h_2|{\cal H}|\bar{M}^0\rangle\equiv \bar{Q}_2$.

From Eq.~(\ref{ittgeneral}),   $I(h_1,t_a;h_1,t_b)$ is obtained by substitution $(r^a,r^b,\bar{r}^a,\bar{r}^b)=(Q_1,Q_1,\bar{Q}_1,\bar{Q}_1)$, $I(h_1,t_a;h_2,t_b)$ by substitution $(r^a,r^b,\bar{r}^a,\bar{r}^b)=(Q_1,Q_2,\bar{Q}_1,\bar{Q}_2)$, $I(h_2,t_a;h_1,t_b)$ by substitution $(r^a,r^b,\bar{r}^a,\bar{r}^b)=(Q_2,Q_1,\bar{Q}_2,\bar{Q}_1)$ and $I(h_2,t_a;h_2,t_b)$ by substitution $(r^a,r^b,\bar{r}^a,\bar{r}^b)=(Q_2,Q_2,\bar{Q}_2,\bar{Q}_2)$.

In this case,  with $|\psi^1\rangle = |h_1\rangle$ and $|\psi^2\rangle=|h_2\rangle$, the   equal-state asymmetry  is
\begin{equation}
\begin{array}{cl}
A(h_1h_1,h_2h_2;\Delta t)
&\approx \frac{\frac{1}{4}(f_0^{Q_1Q_1}-f_0^{Q_2Q_2})+\sum_{i=1}^4(f_i^{Q_1Q_1}-f_i^{Q_2Q_2})\sigma_i}{\frac{1}{4}(f_0^{Q_1Q_1}+f_0^{Q_2Q_2})+\sum_{i=1}^4(f_i^{Q_1Q_1}+f_i^{Q_2Q_2})\sigma_i},
\end{array}\label{A3}
\end{equation}
while the unequal-state asymmetry is
\begin{equation}
\begin{array}{cl}
A(h_1h_2,h_2h_1;\Delta t)&\approx \frac{\sum_{i=1}^4(f_i^{Q_1 Q_2}-f_i^{Q_2Q_1 })\sigma_i}{\frac{1}{2}f_0^{Q_1 Q_2}+\sum_{i=1}^4(f_i^{Q_1Q_2}+f_i^{Q_2Q_1})\sigma_i}.
\end{array}\label{A4}
\end{equation}
They are obtained from Equations (\ref{eq}) and  (\ref{uneq}), respectively, using the substitution  $r^1= Q_1$, $r^2=Q_2$, $\bar{r}^1=\bar{Q}_1$ and $\bar{r}^2= \bar{Q}_2$.

\subsection{Semileptonic-hadronic Process} \label{semihadr}

For semileptonic-hadronic  processes, consider $|\psi^a\rangle=|l^+\rangle$ or $|l^-\rangle$ while  $|\psi^b\rangle=|h_1\rangle$ or $|h_2\rangle$, or vice versa. So there are eight cases of $(\psi^a,\psi^b)$. From Eq.~(\ref{ittgeneral}),
$I(l^+,t_a;h_1,t_b)$ is obtained by substitution $(r^a,r^b,\bar{r}^a,\bar{r}^b)=(R^+,Q_1,\bar{R}^+,\bar{Q}_1)$,
$I(h_1 ,t_a;l^+,t_b)$ by substitution $(r^a,r^b,\bar{r}^a,\bar{r}^b)=(Q_1,R^+,\bar{Q}_1,\bar{R}^+)$,
$I(l^+,t_a;h_2,t_b)$ by substitution $(r^a,r^b,\bar{r}^a,\bar{r}^b)=(R^+,Q_2,\bar{R}^+,\bar{Q}_2)$,
$I(h_2,t_a;l^+,t_b)$ by substitution $(r^a,r^b,\bar{r}^a,\bar{r}^b)=(Q_2,R^+,\bar{Q}_2,\bar{R}^+)$,
$I(l^-,t_a;h_1,t_b)$ by substitution $(r^a,r^b,\bar{r}^a,\bar{r}^b)=(S^-,Q_1,\bar{S}^-,\bar{Q}_1)$,
$I(h_1,t_a;l^-,t_b)$ by substitution $(r^a,r^b,\bar{r}^a,\bar{r}^b)=(Q_1,S^-,\bar{Q}_1,\bar{S}^-)$,
$I(l^-,t_a;h_2,t_b)$ by substitution $(r^a,r^b,\bar{r}^a,\bar{r}^b)=(S^-,Q_2,\bar{S}^-,\bar{Q}_2)$ and
$I(h_2,t_a;l^-,t_b)$ by substitution $(r^a,r^b,\bar{r}^a,\bar{r}^b)=(Q_2,S^-,\bar{Q}_2,\bar{S}^-)$, respectively.

For these eight different outcomes, one can define $28$ different asymmetries according to (\ref{gen}). They are
$A(l^+h_1,h_1l^+,\Delta t)$, $A(l^+h_1,l^+h_2,\Delta t)$, $A(l^+h_1,h_2l^+,\Delta t)$, $A(l^+h_1,l^-h_1,\Delta t)$, $A(l^+h_1,h_1l^-,\Delta t)$, $A(l^+h_1,l^-h_2,\Delta t)$, $A(l^+h_1,h_2l^-,\Delta t)$, $A(h_1l^+,l^+h_2,\Delta t)$, $A(h_1l^+,h_2l^+,\Delta t)$, $A(h_1l^+,l^-h_1,\Delta t)$, $A(h_1l^+,h_1l^-,\Delta t)$, $A(h_1l^+,l^-h_2,\Delta t)$, $A(h_1l^+,h_2l^-,\Delta t)$, $A(l^+h_2,h_2l^+,\Delta t)$, $A(l^+h_2,l^-h_1,\Delta t)$, $A(l^+h_2,h_1l^-,\Delta t)$, $A(l^+h_2,l^-h_2,\Delta t)$, $A(l^+h_2,h_2l^-,\Delta t)$, $A(h_2l^+,l^-h_1,\Delta t)$, $A(h_2l^+,h_1l^-,\Delta t)$, $A(h_2l^+,l^-h_2,\Delta t)$,
$A(h_2l^+,h_2l^-,\Delta t)$, $A(l^-h_1,h_1l^-,\Delta t)$, $A(l^-h_1,l^-h_2,\Delta t)$, $A(l^-h_1,h_2l^-,\Delta t)$, $A(h_1l^-,l^-h_2,\Delta t)$, $A(h_1l^-,h_2l^-,\Delta t)$ and  $A(l^-h_2,h_2l^-,\Delta t)$.

Among them there are four unequal-state asymmetries of the form of (\ref{uneq}),
\begin{equation}
\begin{array}{rcl}
A(l^+h_1,h_1l^+,\Delta t)&\approx& \frac{\sum_{i=1}^4(f_i^{R^+Q_1}-f_i^{Q_1R^+})\sigma_i}{\frac{1}{2}f_0^{R^+Q_1}+\sum_{i=1}^4(f_i^{R^+Q_1}+f_i^{Q_1R^+})\sigma_i},\\
 A(l^+h_2,h_2l^+,\Delta t)&
\approx & \frac{\sum_{i=1}^4(f_i^{R^+Q_2}-f_i^{Q_2R^+})\sigma_i}{\frac{1}{2}f_0^{R^+Q_2}+\sum_{i=1}^4(f_i^{R^+Q_2}+f_i^{Q_2R^+})\sigma_i},\\
 A(l^-h_1,h_1l^-,\Delta t)&
\approx & \frac{\sum_{i=1}^4(f_i^{S^-Q_1}-f_i^{Q_1S^-})\sigma_i}{\frac{1}{2}f_0^{S^-Q_1}+\sum_{i=1}^4(f_i^{S^-Q_1}+f_i^{Q_1S^-})\sigma_i},\\
 A(l^-h_2,h_2l^-,\Delta t)&
\approx & \frac{\sum_{i=1}^4(f_i^{S^-Q_2}-f_i^{Q_2S^-})\sigma_i}{\frac{1}{2}f_0^{S^-Q_2}+\sum_{i=1}^4(f_i^{S^-Q_2}+f_i^{Q_2S^-})\sigma_i}.
\end{array}
\end{equation}

\section{Determining symmetry violating parameters from decay asymmetries}\label{calculation}

We have discussed asymmetries of different decay modes, from which one can determine  the CP and CPT violating parameters. There are four real numbers in the CP and CPT violating parameters. To derive the expressions of the four violating parameters, we need equal number of decay asymmetries.

Suppose we consider four asymmetries $A_k \equiv A(\psi^1_k\psi^2_k,\psi^3_k\psi^4_k;\Delta t)$, with  $k=1,2,3,4$ represent four different joint decay channels. According to (\ref{index}),
\begin{equation}
A_k = \frac{\frac{1}{4}(f_0^{r_k^1r_k^2+}-f_0^{r_k^3r_k^4})
+\sum_{i=1}^4(f_i^{r_k^1r_k^2}-f_i^{r_k^3r_k^4})\sigma_i}{\frac{1}{4}(
f_0^{r_k^1r_k^2}+f_0^{r_k^3r_k^4})+\sum_{i=1}^4(f_i^{r_k^1r_k^2}+f_i^{r_k^3r_k^4})\sigma_i}.
\label{k4}
\end{equation}
Defining
\begin{equation}
\begin{array}{rcl}
a_k&\equiv&\frac{1}{4}[(1-A_k)f_0^{r_k^1r_k^2}-(1+A_k)f_0^{r_k^3r_k^4}],\\
K_{ki}&\equiv&(A_k-1)f_i^{r_k^1r_k^2}+(A_k+1)f_i^{r_k^3r_k^4},
\end{array}
\label{aKx}
\end{equation}
we can rewrite the four equations given by (\ref{k4}) as the following relation between these four asymmetries and the symmetry violating parameters,
\begin{equation}
\left( \begin{array}{c}
a_1\\ a_2\\ a_3\\ a_4
\end{array}\right)=K
\left( \begin{array}{c}
\Re\Delta_M\\ \Im\Delta_M\\ \Re\epsilon_M\\ \Im\epsilon_M
\end{array}\right).
\label{atopgen}
\end{equation}
Hence the CP and CPT symmetry violating parameters can be determined  as
\begin{equation}
\left( \begin{array}{c}
\Re\Delta_M\\ \Im\Delta_M\\ \Re\epsilon_M\\ \Im\epsilon_M
\end{array}\right)=K^{-1}
\left( \begin{array}{c}
a_1\\ a_2\\ a_3\\ a_4
\end{array}\right),
\label{universal}
\end{equation}
where $K^{-1}$ is the inverse matrix of $K$. This provides a general relation between symmetry violating parameters and four arbitrarily chosen decay  asymmetries.

A good choice is to  use the equal-state and unequal-state asymmetries defined for semileptonic-semileptonic processes and hadronic-hadronic  processes. That is, we make the substitutions
\begin{equation}
\begin{array}{rcl}
A_1&=& A(l^+l^+,l^-l^-;\Delta t)\\
A_2&=& A(l^+l^-,l^-l^+;\Delta t)\\
A_3&=& A(h_1h_1,h_2h_2;\Delta t)\\
A_4&=& A(h_1h_2,h_2h_1;\Delta t).
\end{array}
\end{equation}
Then
\begin{equation}
\begin{array}{cl}
a_1\equiv&\frac{1}{4}[(1-A_1)f_0^{R^+R^+}-(1+A_1)f_0^{S^-S^-}],\\
a_2\equiv&-\frac{1}{2}A_2 f_0^{R^+S^-},\\
a_3\equiv&\frac{1}{4}[(1-A_3)f_0^{Q_1Q_1}-(1+A_3)f_0^{Q_2Q_2}],\\
a_4\equiv&-\frac{1}{2}A_4 f_0^{Q_1Q_2},
\end{array}
\label{aai}
\end{equation}
while  the matrix elements of $K$ are given by
\begin{equation}
\begin{split}
K_{1i} \equiv &(A_1-1)f_i^{R^+R^+}+(A_1+1)f_i^{S^-S^-},\\
K_{2i} \equiv &(A_2-1)f_i^{R^+S^-}+(A_2+1)f_i^{S^-R^+},\\
K_{3i} \equiv &(A_3-1)f_i^{Q_1Q_1}+(A_3+1)f_i^{Q_2Q_2},\\
K_{4i} \equiv &(A_4-1)f_i^{Q_1Q_2}+(A_4+1)f_i^{Q_2Q_1},\\
\end{split}
\label{K}
\end{equation}
with $i=1,2,3,4$ and $f^{r^ar^b}_i\equiv f^{r^ar^b}_i(\Delta t)$.

One can also use  some of the asymmetries defined for the  semileptonic-hadronic decay processes.  For example, a convenient choice is to use the four unequal-state asymmetries in  the  semileptonic-hadronic decay processes. Hence one makes the substitutions
\begin{equation}
\begin{array}{rcl}
A_1&=& A(l^+h_1,h_1l^+,\Delta t),\\
A_2&=& A(l^+h_2,h_2l^+,\Delta t),\\
A_3&=& A(l^-h_1,h_1l^-,\Delta t),\\
A_4&=& A(l^-h_2,h_2l^-,\Delta t).
\end{array}
\end{equation}
Then
\begin{equation}
\begin{array}{cl}
a_1\equiv&-\frac{1}{2}A_5f_0^{R^+Q_1},\\
a_2\equiv&-\frac{1}{2}A_6f_0^{R^+\bar{Q}_2},\\
a_3\equiv&-\frac{1}{2}A_7f_0^{S^-Q_1},\\
a_4\equiv&-\frac{1}{2}A_8f_0^{S^-\bar{Q}_2},
\end{array}
\label{a5}
\end{equation}
while the matrix elements of $K$ are given by
\begin{equation}
\begin{split}
K_{1i} = &(A_1-1)f_i^{R^+Q_1}+(A_1+1)f_i^{Q_1R^+},\\
K_{2i} = &(A_2-1)f_i^{R^+Q_2}+(A_2+1)f_i^{Q_2R^+},\\
K_{3i} = &(A_3-1)f_i^{S^-Q_1}+(A_3+1)f_i^{Q_1S^-},\\
K_{4i} = &(A_4-1)f_i^{S^-\bar{Q}_2}+(A_4+1)f_i^{\bar{Q}_2S^-},
\end{split}
\label{K1}
\end{equation}
with $i=1,2,3,4$.

\section{Some theorems   concerning  decay asymmetries and CP and CPT violations  \label{discussions} }

First consider the following situation of equal-time joint decays.  If we exchange $\psi^a$ and $\psi^b$, then $r^a$ and $r^b$ are exchanged, thus $\Theta$ and $\Lambda$ remain  unchanged while $\Xi$ and $\Phi$ are exchanged, consequently, (\ref{deltat}) indicates that $I'(\psi^a,\psi^b,\Delta t=0)=I'(\psi^b,\psi^a,\Delta t=0)$, and thus $A(\psi^a\psi^b,\psi^b\psi^a;\Delta t=0)=0$.

{\bf Theorem 1:} Consider joint decays of $|\Psi_+\rangle$.  For $\Delta t=0$, any unequal-state asymmetry $A(\psi^a\psi^b,\psi^b\psi^a;\Delta t=0)$ always vanishes no matter whether there is CP or CPT violation.

The same conclusion is also valid for $|\Psi_-\rangle$, and has been shown previously for the special cases of  joint decays to flavor eigenstates  and joint decays to CP eigenstates~\cite{yu2}. In the following we show that it is valid for any equal-time unequal-state asymmetry $A(\psi^a\psi^b,\psi^b\psi^a;\Delta t=0)$. From (\ref{minuspsitt}), we obtain for $C=-1$ state,
\begin{equation}
\begin{split}
I(\psi^a,t_a;\psi^b,t_b)=&\big|\langle \psi^a \psi^b|{\cal H}|\psi(t_a,t_b)\rangle\big|^2\\
=&\frac{1}{8}[e^{-(\Gamma_St_a+\Gamma_Lt_b)}|\theta|^2
-2e^{-\Gamma(t_a+t_b)}\Re(\theta^*\lambda e^{i\Delta m\Delta t})+e^{-(\Gamma_Lt_a+\Gamma_St_b)}|\lambda|^2],
\end{split}
\label{decayrate}
\end{equation}
where
\begin{equation}
\begin{split}
\theta&\equiv\eta_2r^ar^b-(1-\xi)r^a\bar{r}^b+(1+\xi)\bar{r}^ar^b-\eta_1\bar{r}^a\bar{r}^b,\\
\lambda&\equiv\eta_2r^ar^b+(1+\xi)r^a\bar{r}^b-
(1-\xi)\bar{r}^ar^b-\eta_1\bar{r}^a\bar{r}^b.
\end{split}
\end{equation}
Obviously when $\psi^a$ and $\psi^b$ are exchanged, so do $\theta$ and $\lambda$. Consequently, when   $t_a=t_b=t$,  $I(\psi^a,t;\psi^b,t)=I(\psi^b,t;\psi^a,t)$, which implies that any equal-time  unequal-state asymmetry $A(\psi^a\psi^b,\psi^b\psi^a;\Delta t=0)$ is zero.

{\bf Theorem 2:} Consider joint decays of $|\Psi_-\rangle$.  For $\Delta t=0$, any unequal-state asymmetry $A(\psi^a\psi^b,\psi^b\psi^a;\Delta t=0)$ always vanishes no matter whether there is CP or CPT violation.

\subsection{Semileptonic-semileptonic Processes}

(i) If  CP is conserved indirectly, then $\epsilon_M=\Delta_M=0$, thus $\Theta=(r^a+\bar{r}^a)(r^b+\bar{r}^b),$ $\Lambda=(r^a-\bar{r}^a)(r^b-\bar{r}^b)$, $\Xi=\Phi=0$.  Consequently, without making any approximation, we obtain exactly
\begin{equation}
\begin{array}{cl}
A(l^+l^+,l^-l^-;\Delta t) &=\frac{f_0^{R^+R^+}(\Delta t)-f_0^{S^-S^-}(\Delta t)}{f_0^{R^+R^+}(\Delta t)+f_0^{S^-S^-}(\Delta t)},\\
A(l^+l^-,l^-l^+;\Delta t)  &=0.\\
\end{array}
\label{a12}
\end{equation}
Any deviation from these two equalities  means indirect CP violation. Especially, a nonvanishing value of unequal-flavor asymmetry $A(l^+l^-,l^-l^+;\Delta t) $ is a signature of indirect CP violation.

{\bf Theorem 3: }   Consider joint decays of $|\Psi_+\rangle$.  If the unequal-flavor asymmetry $A(l^+l^-,l^-l^+;\Delta t) $ is nonzero, then CP must be violated indirectly.

(ii) If CP is conserved directly, then ${R^+}={\bar{S}^-}$ and $ {S^-}={\bar{R}^+}$, consequently
 \begin{equation}
 \begin{split}
  f^{R^+R^+}_0&=f^{S^-S^-}_0,\\
  f^{R^+R^+}_i&=-f^{S^-S^-}_i,\\
  f^{R^+S^-}_j&=f^{S^-R^+}_j=0,\\
  f^{R^+S^-}_0&=f^{S^-R^+}_0,
  \end{split}
  \label{fi}
  \end{equation}
where $i=1,2,3,4$ and $j=3,4$. Then
\begin{equation}
\begin{array}{rcl}
A(l^+l^+,l^-l^-;\Delta t)&\approx& \frac{4\sum_{i=1}^4f_i^{R^+R^+}\sigma_i}{f_0^{R^+R^+}},\\
A(l^+l^-,l^-l^+;\Delta t) &\approx& \frac{\sum_{i=1}^2(f_i^{R^+S^-}-f_i^{S^-R^+})\sigma_i}{\frac{1}{2}f_0^{R^+S^-}+\sum_{i=1}^2(f_i^{R^+S^-}+f_i^{S^-R^+})\sigma_i},
\end{array}
\end{equation}
which says that $A(l^+l^-,l^-l^+;\Delta t) $ does not depend on $\epsilon_M$   up to its first order. Furthermore, if CPT is also assumed to be conserved indirectly, i.e. $\Delta_M=0$, then $A(l^+l^-,l^-l^+;\Delta t)  \sim O(\epsilon_M^2)$.

{\bf Theorem 4: }   Consider joint decays of $|\Psi_+\rangle$.  If  the unequal-flavor asymmetry $A(l^+l^-,l^-l^+;\Delta t)$ depends on the first order of $\epsilon_M$, then CP must be violated directly.

(iii) If the semileptonic decays respect  $\Delta {\cal F}=\Delta Q$ rule, where ${\cal F}$ is the flavor quantum number and $Q$ is charge number,  then ${\bar{R}^+}={S^-}=0$. Consequently
\begin{equation}
\begin{array}{cl}
f^{R^+R^+}_0&=\big|\frac{R^+}{\bar{S}^-}\big|^4f^{S^-S^-}_0,\\
f^{R^+R^+}_i&=-\big|\frac{R^+}{\bar{S}^-}\big|^4 f^{S^-S^-}_i,\\
f^{R^+S^-}_{0}&=f^{S^-R^+}_{0},\\
f^{R^+S^-}_j&=f^{S^-R^+}_j=0,
\end{array}
\label{fi2}
\end{equation}
with $i=1,2,3,4$, $j=3,4$.  Then
\begin{equation}
\begin{array}{rcl}
A(l^+l^+,l^-l^-;\Delta t)&\approx&\frac{\frac{1}{4}(1-|\frac{R^+}{S^-}|^4)f_0^{S^-S^-}+\sum_{i=1}^4(1+|\frac{R^+}{S^-}|^4)f^{S^-S^-}_i\sigma_i}{\frac{1}{4}(1+|\frac{R^+}{S^-}|^4)f_0^{S^-S^-}+\sum_{i=1}^4(1-|\frac{R^+}{S^-}|^4)f^{S^-S^-}_i\sigma_i},\\
A(l^+l^-,l^-l^+;\Delta t) &\approx&\frac{\sum_{i=1}^2(f^{R^+S^-}_i-f^{S^-R^+}_i)\sigma_i}{\frac{1}{2}f^{R^+S^-}_0+\sum_{i=1}^2(f^{R^+S^-}_i+f^{S^-R^+}_i)\sigma_i},
\end{array}
\end{equation}
that is, $A(l^+l^-,l^-l^+;\Delta t) $ does not depend on $\epsilon_M$   up to its first order in this situation.

{\bf Theorem 5:}  Consider joint decays of $|\Psi_+\rangle$.   If the unequal-flavor asymmetry $A(l^+l^-,l^-l^+;\Delta t) $ depends on the first order of $\epsilon_M$, then  $\Delta {\cal F}=\Delta Q$ rule must be violated.

\subsection{Hadronic-hadronic Processes}

(i) For the hadronic-hadronic  processes, first we consider the situation of $|h_1\rangle=CP|h_2\rangle$, that is,  $|h_1\rangle$ and $|h_2\rangle$ are mutual CP conjugates. For example,  $B^0\bar{B}^0\rightarrow D^+K^-D^-K^+, \pi^+D_S^-\pi^-D_S^+$. If CP is conserved directly, then   ${Q_1}={\bar{Q}_2}$ and ${Q_2}={\bar{Q}_1}$, one can  obtain
\begin{equation}
\begin{split}
  f^{Q_1Q_1}_0&=f^{Q_2Q_2}_0,\\
  f^{Q_1Q_1}_i&=-f^{Q_2Q_2}_i,\\
  f^{Q_1Q_2}_{0}&=f^{Q_2Q_1}_{0},\\
  f^{Q_1Q_2}_j&=f^{Q_2Q_1}_j=0,
  \end{split}
  \label{hadr}
\end{equation}
where $i=1,2,3,4$, $j=3,4$. It is obtained that
\begin{equation}
\begin{array}{rcl}
A(h_1h_1,h_2h_2;\Delta t)& \approx &\frac{4\sum_{i=1}^4 f_i^{Q_1Q_1}\sigma_i}{f_0^{Q_1Q_1}},\\
A(h_1h_2,h_2h_1;\Delta t)&\approx &\frac{\sum_{i=1}^2(f^{Q_1Q_2}_i-f^{Q_2Q_1}_i)\sigma_i}{
\frac{1}{2}f^{Q_1Q_2}_0+
\sum_{i=1}^2(f^{Q_1Q_2}_i+f^{Q_2Q_1}_i)\sigma_i}.
\end{array}
\end{equation}
So $A(h_1h_2,h_2h_1;\Delta t)$ does not depend on $\epsilon_M$  up to its first order. Furthermore, if CPT is also assumed to be conserved indirectly, then $\Delta_M=0$, consequently  $A(h_1h_2,h_2h_1;\Delta t)\sim O(\epsilon_M^2)$. One can see the similarity between  $A(h_1h_1,h_2h_2;\Delta t)$ and  $A(l^+l^-,l^-l^+;\Delta t)$, and that between $A(h_1h_2,h_2h_1;\Delta t)$  and $A(l^+l^-,l^-l^+;\Delta t)$.

{\bf Theorem 6:}  Consider joint decays of $|\Psi_+\rangle$.   Suppose  $|h_1\rangle$ and $|h_2\rangle$ are CP conjugates.  If $A(h_1h_1,h_2h_2;\Delta t)$  depends on the first order   of   $\epsilon_M$, then CP must be violated directly.

(ii) Consider the  situation of $CP|h_1\rangle=|h_1\rangle$ and  $CP|h_2\rangle=-|h_2\rangle$, that is, $|h_1\rangle$ and  $|h_2\rangle$ are CP eigenstates with eigenvalues $1$ and $-1$, respectively.
From the expression (\ref{psi+-}) of $|\Psi_+\rangle$ in terms of CP eigenstates,  it is immediately seen that with $\Delta t =0$,  if CP is conserved both directly and indirectly, then  the decay products of the two particle should always be CP eigenstates with an equal eigenvalue, hence $I(h_1,t_a;h_2,t_b) = I(h_2,t_a;h_1,t_b)=0$.

{\bf Theorem 7:}  Consider joint decays of $|\Psi_+\rangle$.   Suppose  $|h_1\rangle$ and  $|h_2\rangle$ are CP eigenstates with eigenvalues $1$ and $-1$, respectively.  The deviation of $I(h_1,t_a;h_2,t_a)$ or $I(h_2,t_a;h_1,t_a)$ from zero  implies CP violation, direct or indirect or both.

In more quantitative details, let us first assume that  CP is conserved directly, then the decay amplitudes satisfy $Q_1=\bar{Q}_1$ and $Q_2=-\bar{Q}_2$, therefore for  $|\psi^a\rangle =|h_1\rangle$ while $|\psi^b\rangle =|h_2\rangle$,  $I(h_1,t_a;h_2,t_b)$ is given by (\ref{ittgeneral}) with $
\Theta=[\eta_2-\eta_1-(\eta_2+\eta_1)\xi]Q_1Q_2$, $
\Xi=(\eta_1+\eta_2+2)\xi Q_1Q_2$, $
\Phi=(\eta_1+\eta_2-2)\xi Q_1Q_2$, $
\Lambda=[\eta_1-\eta_2-(\eta_2+\eta_1)\xi]Q_1Q_2$.
Similarly if $|\psi^a\rangle =|h_2\rangle$ while $|\psi^b\rangle =|h_1\rangle$, then $I(h_2,t_a;h_1,t_b)$ is given by (\ref{ittgeneral}) with$
\Theta=[\eta_2-\eta_1-(\eta_2+\eta_1)\xi]Q_1Q_2,$ $
\Xi=(\eta_1+\eta_2-2)\xi Q_1Q_2$, $
\Phi= (\eta_1+\eta_2+2)\xi Q_1Q_2$, $
\Lambda=[\eta_1-\eta_2-(\eta_2+\eta_1)\xi]Q_1Q_2$.
Consequently, $I(h_1,t_a;h_2,t_b)$ and $I(h_2,t_a;h_1,t_b)$ are of the order of $O(\Delta_M^2)$ and $O(\epsilon_M^2)$. Moreover, if CP is also indirectly conserved, then $\xi=0$, thus $\Theta=\Lambda$ while $\Xi=\Phi$, consequently $I(h_1,t_a;h_2,t_a)= I(h_2,t_a;h_1,t_b)=0$, confirming  Theorem 7.

{\bf Theorem 8:}    Consider joint decays of $|\Psi_+\rangle$.   Suppose  $|h_1\rangle$ and  $|h_2\rangle$ are CP eigenstates with eigenvalues $1$ and $-1$, respectively.  If  $I(h_1,t_a;h_2,t_b)$ and $I(h_2,t_a;h_1,t_b)$ are order of the order of $O(\Delta_M)$ and $O(\epsilon_M)$, then CP is violated directly.

On the other hand, if we first assume that CP is conserved indirectly,  then $\xi=0$ and $\eta_1=\eta_2=1$,  thus   $|M_+(t)\rangle=e^{-i\lambda_St}|M_+\rangle$, $|M_-(t)\rangle=e^{-i\lambda_Lt}|M_-\rangle$. Consequently
\begin{equation}
\begin{split}
|\psi(t_a,t_b)\rangle=\frac{1}{\sqrt{2}}
[e^{-i\lambda_S(t_a+t_b)}|M_+\rangle|M_+\rangle
-e^{-i\lambda_L(t_a+t_b)}|M_-\rangle|M_-\rangle],
\end{split}
\end{equation}
which implies $I(h_1,t_a;h_2,t_b)=I(h_2,t_a;h_1,t_b)=
|e^{-i\Lambda_S (t_a+t_b)} (Q_1+\bar{Q}_1)(Q_2+\bar{Q}_2)- e^{-i\Lambda_L (t_a+t_b)} (Q_1-\bar{Q}_1)(Q_2-\bar{Q}_2)
|^2$. Moreover,  if CP is also conserved directly, then $Q_1= \bar{Q}_1$, $Q_2= -\bar{Q}_2$, consequently $I(h_1,t_a;h_2,t_b)=I(h_2,t_a;h_1,t_b)=0$, again confirming Theorem 7.

\subsection{Semileptonic-hadronic Process}

For semileptonic-hadronic processes, here we consider  some of the asymmetries, for which the assumption of CP conservation can lead to relatively simple results.

(i) Consider the case that the hadronic decay products satisfy $CP|h_1\rangle=|h_2\rangle$. If  CP is conserved directly, then $R^+=\bar{S}^-$, $\bar{R}^+=S^-$, $Q_1=\bar{Q}_2$ and $Q_2=\bar{Q}_1$. Consequently
\begin{equation}
\begin{split}
f^{R^+Q_1}_0&=f^{S^-Q_2}_0,\\
f^{R^+Q_1}_i&=-f^{S^-Q_2}_i,\\
f^{R^+Q_2}_0&=f^{S^-Q_1}_0,\\
f^{R^+Q_2}_i&=-f^{S^-Q_1}_i,
\label{f56}
\end{split}
\end{equation}
with $i=1,2,3,4$. So to the order of $O(\Delta_M)$ and $O(\epsilon_M)$, we have
\begin{equation}
\begin{array}{cl}
A(l^+h_1,l^-h_2;\Delta t)&\approx 4(\frac{f^{R^+Q_1}_1}{f^{R^+Q_1}_0}\Re\epsilon_M+
\frac{f^{R^+Q_1}_2}{f^{R^+Q_1}_0}\Im\epsilon_M+
\frac{f^{R^+Q_1}_3}{f^{R^+Q_1}_0}\Re\Delta_M+
\frac{f^{R^+Q_1}_4}{f^{R^+Q_1}_0}\Im\Delta_M),\\
A(l^+h_2,l^-h_1;\Delta t)&\approx 4(\frac{f^{R^+Q_2}_1}{f^{R^+Q_2}_0}\Re\epsilon_M+
\frac{f^{R^+Q_2}_2}{f^{R^+Q_2}_0}\Im\epsilon_M+
\frac{f^{R^+Q_2}_3}{f^{R^+Q_2}_0}
\Re\Delta_M+\frac{f^{R^+Q_2}_4}{f^{R^+Q_2}_0}\Im\Delta_M).
\end{array}
\label{A56}
\end{equation}

(ii) Consider the case that the hadronic decay products are CP eigenstates, which satisfy $CP|h_1\rangle=|h_1\rangle$ and $CP|h_2\rangle=-|h_2\rangle$. If CP is conserved directly, then $R^+=\bar{S}^-$, $\bar{R}^+=S^-$, $Q_1=\bar{Q}_1$ and $Q_2=-\bar{Q}_2$. Consequently,
\begin{equation}
\begin{split}
f^{R^+Q_k}_0&=f^{S^-Q_k}_0=f^{Q_kS^-}_0,\\
f^{R^+Q_k}_i&=-f^{S^-Q_k}_i=-f^{Q_kS^-}_i,\\
\end{split}
\label{f78}
\end{equation}
with $i=1,2,3,4$ and $k=1,2$. So up to the order of $O(\Delta_M)$ and $O(\epsilon_M)$,  we have
\begin{equation}\begin{split}
A(l^+h_k,l^-h_k;\Delta t)&\approx A(l^+h_k,h_kl^-;\Delta t) \\
& \approx 4(
\frac{f^{R^+Q_k}_1}{f^{R^+Q_k}_0}\Re\epsilon_M+\frac{f^{R^+Q_k}_2}{
f^{R^+Q_k}_0}\Im\epsilon_M+\frac{f^{R^+Q_k}_3}{f^{R^+Q_k}_0}\Re\Delta_M+
\frac{f^{R^+Q_k}_4}{f^{R^+Q_k}_0}\Im\Delta_M),\end{split}
\label{A78}
\end{equation}
with $k=1,2$.

{\bf Theorem 9:}   Consider the semileptonic-hadronic decay  asymmetries $A(l^+h_k,h_kl^-;\Delta t)$ and $A(l^+h_k,l^-h_k;\Delta t)$   of $|\Psi_+\rangle$, $(k=1,2)$.    Suppose  $|h_1\rangle$ and $|h_2\rangle$ are CP eigenstates with eigenvalues $1$ and $-1$, respectively. If $A(l^+h_k,h_kl^-;\Delta t) \neq A(l^+h_k,l^-h_k;\Delta t)$ even in the first order of CP and CPT violating parameters, then CP must be violated directly.

\section{summary \label{summary}}

To summarize, we have studied  $C=+1$ entangled  state $|\Psi_+\rangle$ of pseudoscalar neutral meson pairs. We have calculated  various  integrated joint decay rates of semileptonic-semileptonic processes, hadronic-hadronic processes  and semileptonic-hadronic  processes, as well as experimentally observable  asymmetries defined for them, including equal-state asymmetries, unequal-state asymmetries and more general ones, which are functions of the CP and CPT violating parameters.  Any four of these asymmetries can be used to determine   the real and imaginary parts of the indirect  symmetry violating parameters $\epsilon_M$ and $\Delta_M$.  For example, one can choose the equal-state and unequal-state asymmetries in semileptonic-semileptonic and hadronic-hadronic decays.  Alternatively, one can choose four asymmetries in semileptonic-hadronic decays.   The coefficients in these equations depend on whether CP is violated directly or whether $\Delta {\cal F} =\Delta Q$ rule is violated indirectly or directly. Through these relations we can examine whether various symmetries or rules are violated, and determine the symmetry violating parameters. Also note that these relations are for a given $\Delta t$, hence by using various different values, one can obtain the quantities in many times, and make averages.

We also make some simple statements concerning the joint decays of $|\Psi_+\rangle$  presented as theorems, as the following.  If the unequal-flavor asymmetry $A(l^+l^-,l^-l^+;\Delta t) $ is nonzero, then CP must be violated indirectly. If $A(l^+l^-,l^-l^+;\Delta t)$ depends on the first order of $\epsilon_M$, then CP must be violated directly, and $\Delta {\cal F}=\Delta Q$ rule is violated.     If $A(h_1h_1,h_2h_2;\Delta t)$  for CP conjugates $|h_1\rangle$ and $|h_2\rangle$ depends on the first order   of   $\epsilon_M$, then CP is violated directly. For   CP eigenstates   $|h_1\rangle$ and  $|h_2\rangle$  with eigenvalues $1$ and $-1$, respectively, the deviation of $I(h_1,t_a;h_2,t_b)$ or $I(h_2,t_a;h_1,t_b)$ from zero  implies CP violation.
For CP eigenstates  $|h_1\rangle$ and $|h_2\rangle$    with eigenvalues $1$ and $-1$, respectively,  consider the semileptonic-hadronic decay  asymmetries $A(l^+h_k,h_kl^-;\Delta t)$ and $A(l^+h_k,l^-h_k;\Delta t)$, $(k=1,2)$. If $A(l^+h_k,h_kl^-;\Delta t) \neq A(l^+h_k,l^-h_k;\Delta t)$ even in the first order of CP and CPT violating parameters, then CP must be violated directly.

The uses of $|\Psi_+\rangle$ and $|\Psi_-\rangle$ well complement each other. Two outstanding examples are as follows.  In  flavor basis, their relative phases between $|M^0\rangle|\bar{M}^0\rangle$ and $|\bar{M}^0\rangle|M^0\rangle$ are opposite. Consequently,   CP must be violated indirectly if the unequal-flavor asymmetry in $|\Psi_+\rangle$ is nonzero,   while  if the equal-flavor asymmetry in $|\Psi_-\rangle$is nonzero~\cite{yu2}. On the other hand, in CP basis, $|\Psi_+\rangle$ is a superposition of equal-CP terms $|M_+\rangle|M_+\rangle$ and $|M_-\rangle|M_-\rangle$, while $|\Psi_-\rangle$ is a superposition of unequal-CP terms $|M_-\rangle|M_+\rangle$ and $|M_+\rangle|M_-\rangle$. Consequently, CP must be violated if any unequal-CP joint decay rate of $|\Psi_+\rangle$  is nonzero, while if any equal-CP joint decay rate of $|\Psi_-\rangle$  is nonzero~\cite{yu2}. Besides,  $|\Psi_+\rangle$ and $|\Psi_-\rangle$  also share some common phenomena. For example, for both   $|\Psi_+\rangle$ and $|\Psi_-\rangle$, any equal-time unequal-state asymmetry $A(\psi^a\psi^b,\psi^b\psi^a;\Delta t=0)$   must always vanish no matter whether CP and CPT are violated.

We hope the present study of $|\Psi_+\rangle$ motivates its use in studying CP and CPT violations. Among various reasons,  note the availability of  $|\Psi_+\rangle$ is available in an energy range just  above the  $\Upsilon(4\textbf{S})$ resonance with 100\% branch ratio~\cite{akerib}.

\appendix

\section{Detailed expressions of $\tilde{f}_i^{r^ar^b\bar{r}^a\bar{r}^b}$ and  $f_i^{r^ar^b\bar{r}^a\bar{r}^b}$ }

\begin{equation*}
\begin{array}{cl}
\tilde{f}_0^{r^ar^b}(t_a,t_b)=&|(r^a+\bar{r}^a)(r^b+\bar{r}^b)|^2e^{-\Gamma_S(t_a+t_b)}+|(r^a-\bar{r}^a)(r^b-\bar{r}^b)|^2e^{-\Gamma_L(t_a+t_b)}\\
&-2e^{-\Gamma(t_a+t_b)}\Re\big[(r^a+\bar{r}^a)^*(r^b+\bar{r}^b)^*(r^a-\bar{r}^a)(r^b-\bar{r}^b)e^{-i\Delta m(t_a+t_b)}\big],
\end{array}
\end{equation*}
\begin{equation*}
\begin{array}{cl}
f_0^{r^ar^b}(\Delta t)=&e^{-\Gamma\Delta t}\big\{\frac{|(r^a+\bar{r}^a)(r^b+\bar{r}^b)|^2}{2\Gamma_S}e^{\frac{\Delta\Gamma\Delta t}{2}}+\frac{|(r^a-\bar{r}^a)(r^b-\bar{r}^b)|^2}{2\Gamma_L}e^{-\frac{\Delta\Gamma\Delta t}{2}}\\
&-\Re\big[\frac{(r^a+\bar{r}^a)^*(r^b+\bar{r}^b)^*(r^a-\bar{r}^a)(r^b-\bar{r}^b)}{(\Gamma+i\Delta m)}e^{-i\Delta m\Delta t}\big]\big\},
\end{array}
\end{equation*}
\begin{equation*}
\begin{array}{cl}
\tilde{f}_1^{r^ar^b}(t_a,t_b)=&-e^{-\Gamma_S(t_a+t_b)}\Re\big[(r^ar^b-\bar{r}^a\bar{r}^b)^*(r^a+\bar{r}^a)(r^b+\bar{r}^b)\big]\\
&+e^{-\Gamma_L(t_a+t_b)}\Re\big[(r^ar^b-\bar{r}^a\bar{r}^b)^*(r^a-\bar{r}^a)(r^b-\bar{r}^b)\big]\\
&-e^{-\Gamma(t_a+t_b)}\Re\big[(r^ar^b-\bar{r}^a\bar{r}^b)^*(r^a+\bar{r}^a)(r^b+\bar{r}^b)e^{i\Delta m(t_a+t_b)}\big]\\
&+e^{-\Gamma(t_a+t_b)}\Re\big[(r^ar^b-\bar{r}^a\bar{r}^b)^*(r^a-\bar{r}^a)(r^b-\bar{r}^b)e^{-i\Delta m(t_a+t_b)}\big]\\
&+e^{-(\Gamma t_a+\Gamma_S t_b)}|r^b+\bar{r}^b|^2\Re\big[(r^a+\bar{r}^a)^*(r^a-\bar{r}^a)e^{-i\Delta mt_a}\big]\\
&+e^{-(\Gamma_St_a+\Gamma t_b)}|r^a+\bar{r}^a|^2\Re\big[(r^b+\bar{r}^b)^*(r^b-\bar{r}^b)e^{-i\Delta mt_b}\big]\\
&-e^{-(\Gamma t_a+\Gamma_Lt_b)}|r^b-\bar{r}^b|^2\Re\big[(r^a-\bar{r}^a)^*(r^a+\bar{r}^a)e^{i\Delta mt_a}\big]\\
&-e^{-(\Gamma_Lt_a+\Gamma t_b)}|r^a-\bar{r}^a|^2\Re\big[(r^b-\bar{r}^b)^*(r^b+\bar{r}^b)e^{i\Delta mt_b}\big],
\end{array}
\end{equation*}
\begin{equation*}
\begin{array}{cl}
f_1^{r^ar^b}(\Delta t)=&e^{-\Gamma\Delta t}\Big\{-\frac{\Re[(r^ar^b-\bar{r}^a\bar{r}^b)^*(r^a+\bar{r}^a)(r^b+\bar{r}^b)]}{2\Gamma_S}e^{\frac{\Delta\Gamma\Delta t}{2}}+\frac{\Re[(r^ar^b-\bar{r}^a\bar{r}^b)^*(r^a-\bar{r}^a)(r^b-\bar{r}^b)]}{2\Gamma_L}e^{-\frac{\Delta\Gamma\Delta t}{2}}\\
&-\Re\big[\frac{(r^ar^b-\bar{r}^a\bar{r}^b)^*(r^a+\bar{r}^a)(r^b+\bar{r}^b)}{2(\Gamma-i\Delta m)}e^{i\Delta m\Delta t}\big]+\Re\big[\frac{(r^ar^b-\bar{r}^a\bar{r}^b)^*(r^a-\bar{r}^a)(r^b-\bar{r}^b)}{2(\Gamma+i\Delta m)}e^{-i\Delta m\Delta t}\big]\\
&+|r^b+\bar{r}^b|^2\Re\big[\frac{(r^a+\bar{r}^a)^*(r^a-\bar{r}^a)}{\Gamma+\Gamma_S+i\Delta m}\big]e^{-\frac{\Delta\Gamma\Delta t}{2}}+|r^a+\bar{r}^a|^2\Re\big[\frac{(r^b+\bar{r}^b)^*(r^b-\bar{r}^b)}{\Gamma+\Gamma_S+i\Delta m}e^{-i\Delta m\Delta t}\big]\\
&-|r^b-\bar{r}^b|^2\Re\big[\frac{(r^a-\bar{r}^a)^*(r^a+\bar{r}^a)}{\Gamma+\Gamma_L-i\Delta m}\big]e^{\frac{\Delta\Gamma\Delta t}{2}}-|r^a-\bar{r}^a|^2\Re\big[\frac{(r^b-\bar{r}^b)^*(r^b+\bar{r}^b)}{\Gamma+\Gamma_L-i\Delta m}e^{i\Delta m\Delta t}\big]
\Big\},
\end{array}
\end{equation*}
\begin{equation*}
\begin{array}{cl}
\tilde{f}_2^{r^ar^b}(t_a,t_b)=&-e^{-\Gamma_S(t_a+t_b)}\Im\big[(r^ar^b-\bar{r}^a\bar{r}^b)^*(r^a+\bar{r}^a)(r^b+\bar{r}^b)\big]\\
&+e^{-\Gamma_L(t_a+t_b)}\Im\big[(r^ar^b-\bar{r}^a\bar{r}^b)^*(r^a-\bar{r}^a)(r^b-\bar{r}^b)\big]\\
&-e^{-\Gamma(t_a+t_b)}\Im\big[(r^ar^b-\bar{r}^a\bar{r}^b)^*(r^a+\bar{r}^a)(r^b+\bar{r}^b)e^{i\Delta m(t_a+t_b)}\big]\\
&+e^{-\Gamma(t_a+t_b)}\Im\big[(r^ar^b-\bar{r}^a\bar{r}^b)^*(r^a-\bar{r}^a)(r^b-\bar{r}^b)e^{-i\Delta m(t_a+t_b)}\big]\\
&-e^{-(\Gamma t_a+\Gamma_S t_b)}|r^b+\bar{r}^b|^2\Im\big[(r^a+\bar{r}^a)^*(r^a-\bar{r}^a)e^{-i\Delta mt_a}\big]\\
&-e^{-(\Gamma_St_a+\Gamma t_b)}|r^a+\bar{r}^a|^2\Im\big[(r^b+\bar{r}^b)^*(r^b-\bar{r}^b)e^{-i\Delta mt_b}\big]\\
&+e^{-(\Gamma t_a+\Gamma_Lt_b)}|r^b-\bar{r}^b|^2\Im\big[(r^a-\bar{r}^a)^*(r^a+\bar{r}^a)e^{i\Delta mt_a}\big]\\
&+e^{-(\Gamma_Lt_a+\Gamma t_b)}|r^a-\bar{r}^a|^2\Im\big[(r^b-\bar{r}^b)^*(r^b+\bar{r}^b)e^{i\Delta mt_b}\big],
\end{array}
\end{equation*}
\begin{equation*}
\begin{array}{cl}
f_2^{r^ar^b}(\Delta t)=&e^{-\Gamma\Delta t}\Big\{-\frac{\Im[(r^ar^b-\bar{r}^a\bar{r}^b)^*(r^a+\bar{r}^a)(r^b+\bar{r}^b)]}{2\Gamma_S}e^{\frac{\Delta\Gamma\Delta t}{2}}+\frac{\Im[(r^ar^b-\bar{r}^a\bar{r}^b)^*(r^a-\bar{r}^a)(r^b-\bar{r}^b)]}{2\Gamma_L}e^{-\frac{\Delta\Gamma\Delta t}{2}}\\
&-\Im\big[\frac{(r^ar^b-\bar{r}^a\bar{r}^b)^*(r^a+\bar{r}^a)(r^b+\bar{r}^b)}{2(\Gamma-i\Delta m)}e^{i\Delta m\Delta t}\big]+\Im\big[\frac{(r^ar^b-\bar{r}^a\bar{r}^b)^*(r^a-\bar{r}^a)(r^b-\bar{r}^b)}{2(\Gamma+i\Delta m)}e^{-i\Delta m\Delta t}\big]\\
&-|r^b+\bar{r}^b|^2\Im\big[\frac{(r^a+\bar{r}^a)^*(r^a-\bar{r}^a)}{\Gamma+\Gamma_S+i\Delta m}\big]e^{\frac{\Delta\Gamma\Delta t}{2}}-|r^a+\bar{r}^a|^2\Im\big[\frac{(r^b+\bar{r}^b)^*(r^b-\bar{r}^b)}{\Gamma+\Gamma_S+i\Delta m}e^{-i\Delta m\Delta t}\big]\\
&+|r^b-\bar{r}^b|^2\Im\big[\frac{(r^a-\bar{r}^a)^*(r^a+\bar{r}^a)}{\Gamma+\Gamma_L-i\Delta m}\big]e^{-\frac{\Delta\Gamma\Delta t}{2}}+|r^a-\bar{r}^a|^2\Im\big[\frac{(r^b-\bar{r}^b)^*(r^b+\bar{r}^b)}{\Gamma+\Gamma_L-i\Delta m}e^{i\Delta m\Delta t}\big]\Big\},
\end{array}
\end{equation*}
\begin{equation*}
\begin{split}
\tilde{f}_3^{r^ar^b}(t_a,t_b)=&e^{-\Gamma_S(t_a+t_b)}\Re\big[(r^ar^b-\bar{r}^a\bar{r}^b)^*(r^a+\bar{r}^a)(r^b+\bar{r}^b)\big]\\
&+e^{-\Gamma_L(t_a+t_b)}\Re\big[(r^ar^b-\bar{r}^a\bar{r}^b)^*(r^a-\bar{r}^a)(r^b-\bar{r}^b)\big]\\
&-e^{-\Gamma(t_a+t_b)}\Re\big[(r^ar^b-\bar{r}^a\bar{r}^b)^*(r^a+\bar{r}^a)(r^b+\bar{r}^b)e^{i\Delta m(t_a+t_b)}\big]\\
&-e^{-\Gamma(t_a+t_b)}\Re\big[(r^ar^b-\bar{r}^a\bar{r}^b)^*(r^a-\bar{r}^a)(r^b-\bar{r}^b)e^{-i\Delta m(t_a+t_b)}\big],
\end{split}
\end{equation*}
\begin{equation*}
\begin{array}{cl}
f_3^{r^ar^b}(\Delta t)=&e^{-\Gamma\Delta t}\Big\{\frac{\Re[(r^ar^b-\bar{r}^a\bar{r}^b)^*(r^a+\bar{r}^a)(r^b+\bar{r}^b)]}{2\Gamma_S}e^{\frac{\Delta\Gamma\Delta t}{2}}+\frac{\Re[(r^ar^b-\bar{r}^a\bar{r}^b)^*(r^a-\bar{r}^a)(r^b-\bar{r}^b)]}{2\Gamma_L}e^{-\frac{\Delta\Gamma\Delta t}{2}}\\
&-\Re\big[\frac{(r^ar^b-\bar{r}^a\bar{r}^b)^*(r^a+\bar{r}^a)(r^b+\bar{r}^b)}{2(\Gamma-i\Delta m)}e^{i\Delta m\Delta t}\big]-\Re\big[\frac{(r^ar^b-\bar{r}^a\bar{r}^b)^*(r^a-\bar{r}^a)(r^b-\bar{r}^b)}{2(\Gamma+i\Delta m)}e^{-i\Delta m\Delta t}\big]\Big\},
\end{array}
\end{equation*}
\begin{equation*}
\begin{split}
\tilde{f}_4^{r^ar^b}(t_a,t_b)=&e^{-\Gamma_S(t_a+t_b)}\Im\big[(r^ar^b-\bar{r}^a\bar{r}^b)^*(r^a+\bar{r}^a)(r^b+\bar{r}^b)\big]\\
&+e^{-\Gamma_L(t_a+t_b)}\Im\big[(r^ar^b-\bar{r}^a\bar{r}^b)^*(r^a-\bar{r}^a)(r^b-\bar{r}^b)\big]\\
&-e^{-\Gamma(t_a+t_b)}\Im\big[(r^ar^b-\bar{r}^a\bar{r}^b)^*(r^a+\bar{r}^a)(r^b+\bar{r}^b)e^{i\Delta m(t_a+t_b)}\big]\\
&-e^{-\Gamma(t_a+t_b)}\Im\big[(r^ar^b-\bar{r}^a\bar{r}^b)^*(r^a-\bar{r}^a)(r^b-\bar{r}^b)e^{-i\Delta m(t_a+t_b)}\big],
\end{split}
\end{equation*}
\begin{equation*}
\begin{array}{cl}
f_4^{r^ar^b}(\Delta t)=&e^{-\Gamma\Delta t}\Big\{\frac{\Im[(r^ar^b-\bar{r}^a\bar{r}^b)^*(r^a+\bar{r}^a)(r^b+\bar{r}^b)]}{2\Gamma_S}e^{\frac{\Delta\Gamma\Delta t}{2}}+\frac{\Im[(r^ar^b-\bar{r}^a\bar{r}^b)^*(r^a-\bar{r}^a)(r^b-\bar{r}^b)]}{2\Gamma_L}e^{-\frac{\Delta\Gamma\Delta t}{2}}\\
&-\Im\big[\frac{(r^ar^b-\bar{r}^a\bar{r}^b)^*(r^a+\bar{r}^a)(r^b+\bar{r}^b)}{2(\Gamma-i\Delta m)}e^{i\Delta m\Delta t}\big]-\Im\big[\frac{(r^ar^b-\bar{r}^a\bar{r}^b)^*(r^a-\bar{r}^a)(r^b-\bar{r}^b)}{2(\Gamma+i\Delta m)}e^{-i\Delta m\Delta t}\big]
\Big\}.
\end{array}
\end{equation*}

\acknowledgments

This work was supported by the National Science Foundation of China (Grant No. 10875028). We  thank Sheldon  Stone  for drawing our attention to Ref.~\cite{akerib}.

\end{document}